# Landau quantization, Rashba spin-orbit coupling and Zeeman splitting of two-dimensional heavy holes

S.A.Moskalenko[1], I.V.Podlesny[1], E.V.Dumanov[1], M.A.Liberman[2,3], B.V.Novikov[4]

[1]Institute of Applied Physics of the Academy of Sciences of Moldova, Academic Str. 5, Chisinau 2028, Republic of Moldova
[2]Nordita, KTH Royal Institute of Technology and Stockholm University, Roslagstullsbacken 23, 10691 Stockholm, Sweden
[3]Moscow Institute of Physics and Technology, Inststitutskii per. 9, Dolgoprudnyi, Moskovsk. obl.,141700, Russia
[4]Department of Solid State Physics, Institute of Physics, St.Petersburg State University, 1, Ulyanovskaya str., Petrodvorets, 198504 St. Petersburg, Russia

## Abstract

The origin of the g-factor of the two-dimensional (2D) electrons and holes moving in the periodic crystal lattice potential with the perpendicular magnetic and electric fields is discussed. The Pauli equation describing the Landau quantization accompanied by the Rashba spin-orbit coupling (RSOC) and Zeeman splitting (ZS) for 2D heavy holes with nonparabolic dispersion law is solved exactly. The solutions have the form of the pairs of the Landau quantization levels due to the spinor-type wave functions. The energy levels depend on amplitudes of the magnetic and electric fields, on the g-factor $g_h$, and on the parameter of nonparabolicity $C$. The dependences of two energy levels in any pair on the Zeeman parameter $Z_h = g_h \frac{m_h}{4m_0}$, where $m_h$ is the hole effective mass, are nonmonotonous and without intersections. The smallest distance between them at $C=0$ takes place at the value $Z_h = n/2$, where $n$ is the order of the chirality terms determined by the RSOC and is the same for any quantum number of the Landau quantization.

## 1. Introduction

The spin properties of electrons and holes in semiconductors are strongly related with their movement in the frame of the crystal lattice periodic potential. The periodic lattice potential and the energy band structure of the crystal essentially influence the spin states of the charged particles including their g-factor and Zeeman splitting in the presence of the magnetic fields. In spite of the fact that the spin is a relativistic, intrinsic and profound property of the matter, nevertheless it is very sensitive to the particular conditions, in which semiconductor structure was grown. We will remind shortly why it occurs, starting with the relativistic Dirac equation and finishing with the interplay between the Landau quantization, Rashba spin-orbit coupling and Zeeman splitting in the case of the two-dimensional hole gas (2DHG). The important role of the periodic lattice potential in the formation of the spin and of the total angular momentum states is determined by the symmetric properties of the crystals. Brown [1] emphasized that theoretical understanding of the electrons and holes behavior in crystalline potentials is enormously simplified by virtue of the invariance properties of the Hamiltonian under the operations of the space groups. Following Onsager [2], we consider the effective Hamiltonian which depends on the kinetic momentum $\vec{\pi} = \hat{\vec{P}} - \frac{q}{c}\vec{A}(\vec{r})$, instead of the canonical momentum $\hat{\vec{P}}$, where $q$ is the electric charge of the particle and $\vec{A}(\vec{r})$ is the vector potential generating the magnetic field $\vec{B} = \text{rot}\vec{A}$ and obeying to the Maxwell equation $\text{div}\vec{A}(\vec{r}) = 0$ in any gauge. The effective Hamiltonian $\hat{H}_{eff}\left(\hat{\vec{P}} - \frac{q}{c}\vec{A}(\vec{r})\right)$ commutes with any function, which depends on $\left(\hat{\vec{P}} + \frac{q}{c}\vec{A}(\vec{r})\right)$, including the operators of the magnetic translations

$$T(\vec{R}_n) = \exp\left[-\frac{i\vec{R}_n\left(\hat{\vec{P}} + \frac{q}{c}\vec{A}(\vec{r})\right)}{\hbar}\right] \quad (1)$$

Note, that the operators $\left(\hat{\vec{P}} - \frac{q}{c}\vec{A}(\vec{r})\right)$ and $\left(\hat{\vec{P}} + \frac{q}{c}\vec{A}(\vec{r})\right)$ do commute due to $\text{div}\vec{A}(\vec{r}) = 0$.

In [1] it was shown that the operators of the magnetic translations $T(\vec{R}_n)$ can be presented as the products of the pure spatial translation operators $\exp\left(-i\frac{\vec{R}_n \cdot \hat{\vec{P}}}{\hbar}\right)$ and the Paierls phase terms [3] $\exp\left(-\frac{iq}{\hbar c}\vec{R}_n \cdot \vec{A}(\vec{r})\right)$. These two terms commute due to $\text{div}\vec{A} = 0$. The magnetic translation operators do not constitute a group, but only a ray group, because the product of any two of them is not necessarily also a translation operator, which may differ by a phase factor. This leads to the accumulation of the phase factors during successive magnetic translations. For the electron with the electric charge $q = -|e|$ after the three steps closed round, we obtain

$$T(\vec{R}_1)T(\vec{R}_2)T(\vec{R}_3) = T(\vec{R}_1 + \vec{R}_2 + \vec{R}_3)\exp\left\{-\frac{i\vec{\beta}}{2}[\vec{R}_2 \times \vec{R}_3]\right\}\exp\left\{-\frac{i\vec{\beta}}{2}[\vec{R}_1 \times (\vec{R}_2 + \vec{R}_3)]\right\} = e^{-2\pi i\frac{\phi}{\phi_0}},$$

$$\vec{\beta} = \frac{|e|}{\hbar c}\vec{B}, \; \phi = \vec{S}\vec{B}, \; \vec{S} = \frac{[\vec{R}_2 \times \vec{R}_3]}{2}, \; \phi_0 = \frac{2\pi\hbar c}{|e|}, \; \vec{R}_1 + \vec{R}_2 + \vec{R}_3 = 0 \quad (2)$$

Here $\phi$ is the magnetic flux through the surface $\vec{S}$ of the triangle determined by the vectors $\vec{R}_1, \vec{R}_2, \vec{R}_3$, and $\phi_0$ is the magnetic flux quantum. This marvelous result determines the Aharonov–Bohm effect [4].

The theory of the optical magneto-absorption effects in semiconductors was developed by Roth, Lax and Zwerling [5] using the basic concepts developed in [6]. For a simple parabolic conduction and valence bands and the direct transitions allowed in the point $\vec{k} = 0$ of the Brillouin zone, the selection rule in the presence of the magnetic field was established in the form $\Delta n = 0$, where n is the quantum number of the Landau quantization levels. The spacing between the peaks of the band-to-band optical transitions is the sum of the cyclotron frequencies for two bands. In the case of the degenerate valence band the six fold degeneracy includes the spin splitting by the spin-orbit interaction into a four fold and two fold degenerate set. The basis functions for the 3/2 angular momentum constructed from the p-like spatial functions x, y, z and from up $|\alpha\rangle$ and down $|\beta\rangle$ spin functions reads [5]:

$$j_z = 3/2, \quad u_{10} = \frac{1}{\sqrt{2}}(x+iy)|\alpha\rangle,$$

$$j_z = -3/2, \quad u_{20} = \frac{1}{\sqrt{2}}(x-iy)|\beta\rangle,$$

$$j_z = 1/2, \quad u_{30} = \frac{1}{\sqrt{6}}[(x+iy)|\beta\rangle - 2z|\alpha\rangle], \quad (3)$$

$$j_z = -1/2, \quad u_{40} = \frac{1}{\sqrt{6}}[(x-iy)|\alpha\rangle + 2z|\beta\rangle].$$

In the 2D structures due to the size quantization, energy of state of the in-plane wave functions $(x \pm iy)/\sqrt{2}$ differs form the energy of the z function oriented along the growth direction. It makes sense to consider the Bloch electron wave functions $(x \pm iy)/\sqrt{2}|\alpha\rangle$ and $(x \pm iy)/\sqrt{2}|\beta\rangle$, even without their mixing with $z|\alpha\rangle$ and $z|\beta\rangle$ functions. This is important in the case of the two-dimensional hole gas in such structures, as e.g. GaAs quantum wells. Authors of Ref. [7] emphasized that: "Despite the importance of GaAs for fundamental research and technological applications, a detailed study of the effective mass of holes in GaAs 2DHG grown along the high symmetry [001] direction remains to be done. The interpretation of the rapidly increasing number of experiments performed in 2DHGs requires a solid understanding of the physics underlying the effective mass value and its dependence on quantities such as hole density and spin-orbit interaction (SOI) strength."

The fine structure of excitons in type II GaAs/AlAs quantum wells, the electron-hole exchange interaction and the g-factor for the electrons and holes were obtained for GaAs and AlAs layers of different thicknesses [8]. The anisotropy of the electron g-factor is in accordance with the increase of the three fold degeneracy of the AlAs X conduction band minimum by the quantum well potential. The effective heavy-hole g value ~2.5 is much smaller than in the bulk and depends on the GaAs well thickness. Perhaps this is a consequence of the valence band mixing in quantum well structures [8]. The exciton, heavy hole and electron g-factor in type I GaAs/Al$_x$Ga$_{1-x}$As quantum wells were investigated and their dependences on the QW thicknesses were obtained [9]. In the [10] the diamagnetic shift and the Zeeman splitting of the spatially confined polariton modes in an InGaAs quantum well microcavity were investigated in the external magnetic field and different trap diameters ranging from 1 to 10 $\mu$m. The small polariton effective mass enables the observation of the 3D quantization effects for trap diameters of a few micrometers. The lateral confinement of polaritons creates opportunities of realization of future quantum light emitters. The confined polaritons become a source of indistinguishable single photons. A single polariton level in a photonic dot with high confinement potential for polaritons can act as a single photon source [10].

The purpose of the present paper is to increase the knowledge about the 2D heavy-holes taking into account the nonparabolicity of their dispersion law induced by the perpendicular to the layer external electric field giving rise to the Rashba spin-orbit coupling with third order chirality terms. It means that the nondiagonal terms in the Pauli equation have a cubic dependence on the in-plane momentum. The paper is organized as follows. In the section 2 the origin of the g-factors and effective masses of 2D electrons and holes moving in the crystal lattice potential and in the perpendicular magnetic and electric fields is traced beginning with the Dirac equation for the relativistic charged particle with spin 1/2 and finishing with the Pauli equation in the quasi-relativistic approximation. The relation between the g-factor for 2D heavy and light holes is proposed. In section 3 the Pauli equation describing the Landau quantization, the Rashba spin-orbit coupling and the Zeeman splitting for 2D heavy holes with nonparabolic dispersion law induced by the external electric field is solved exactly. The eigenvalues of the energy levels in dependence on the magnetic and electric fields strengths and on the g-factor values were discussed. We conclude in section 4.

## 2. The Pauli equation for the two-dimensional electron in the crystal lattice potential and a perpendicular magnetic field

The Dirac equation for the relativistic particle with spin 1/2 in the quasi-relativistic approximation leads to the Pauli equations as it was shown in the Davydov's textbook [11]. Taking into account the original Dirac equation and following [11], we obtain the Pauli equation for a particle with bare mass $m_0$ and electric charge $q$ moving in the crystal with applied external magnetic field:

$$\varepsilon \hat{\varphi} = \left\{ \frac{\left(\vec{P} - \frac{q}{c}\vec{A}(\vec{r})\right)^2}{2m_0} \hat{I} + V_0(\vec{r})\hat{I} - \frac{q\hbar}{2m_0 c}\vec{\sigma}\vec{B} \right\} \hat{\varphi} \qquad (4)$$

Here $\varepsilon$ is the eigenvalue of the particle energy excluding the rest mass energy $m_0 c^2$, $\hat{\varphi}$ is the spinor-type wave function, $\vec{B} = \text{rot}\vec{A}(\vec{r})$, and $\vec{\sigma}$ is the vector with three components $\hat{\sigma}_x, \hat{\sigma}_y, \hat{\sigma}_z$ and $\hat{I}$ is the unit matrix of the second order. The crystal lattice potential $V_0(\vec{r})$ is a periodic function, which is invariant under the translation on the integer vector $\vec{R}$ of the lattice, enumerating the lattice nodes: $V_0(\vec{\rho} + \vec{R}) = V_0(\vec{\rho})$. This property allows to represent the arbitrary real space vector $\vec{r}$ in the crystal in the form $\vec{r} = \vec{R} + \vec{\rho}$, where $\vec{\rho}$ is a small vector changing inside the unit lattice cell with the volume $v_0 = a_0^3$ and lattice constant $a_0$. Due to this property the electron Bloch wave functions can be represented as the products of the periodic parts $u_{n,k}(\vec{\rho})$ quickly changing inside the electron shells of the atoms situated into the unit lattice cells and depending only on the variable $\vec{\rho}$ and of the envelope functions $\varphi_{n,k}(\vec{R})$ slowly varying on the variable $\vec{R}$ and do not depending practically on the variable $\vec{\rho}$

$$\psi_{n,k}(\vec{r}) = u_{n,k}(\vec{\rho})\varphi_{n,k}(\vec{R}) = \psi_{n,k}(\vec{\rho},\vec{R}) \tag{5}$$

Due to this property the summation on the discrete lattice nodes can be substituted by the integration on the continuum of the values $\vec{R}$ as follows

$$\int d^2\vec{r} = \sum_{\vec{R}} \int_{S_0} d^2\vec{\rho} = \sum_{\vec{R}} a_0^2 \frac{1}{s_0} \int_{S_0} d^2\vec{\rho} = \int d^2\vec{R} \frac{1}{v_0} \int_{v_0} d^3\vec{\rho},$$
$$S_0 = a_0^2,\ v_0 = a_0^3,\ \vec{r} = \vec{R} + \vec{\rho},\ a_0^2 \cong d^2\vec{R} \tag{6}$$

The integration over the surface area $S_0$ of the unit lattice cell is substituted by the integration on its volume $v_0$ because 2D semiconductor structures have at least the thickness $a_0$ of the one atomic layer. Taking into account that $\vec{A}(\vec{r})$ depends linearly on $\vec{r}$, we can write in two-variables ($\vec{\rho}$ and $\vec{R}$) representation

$$\vec{A}(\vec{r}) = \vec{A}(\vec{\rho}) + \vec{A}(\vec{R}); \frac{\partial}{\partial \vec{r}}\psi(\vec{\rho},\vec{R}) = \frac{\partial}{\partial \vec{\rho}}\psi(\vec{\rho},\vec{R}) + \frac{\partial}{\partial \vec{R}}\psi(\vec{\rho},\vec{R})$$
$$\vec{p}_{\vec{r}} = \vec{p}_\rho + \vec{p}_R; \vec{B}(\vec{\rho}) = rot_\rho \vec{A}(\vec{\rho}); \vec{B}(\vec{R}) = rot_{\vec{R}} \vec{A}(\vec{R}) \tag{7}$$

The magnetic field $\vec{B}(\vec{\rho})$ and $\vec{B}(\vec{R})$ are acting in different points of the real space such as in the electron shell of the atoms situated inside the unit lattice cell giving rise to their Zeeman effects as well as outside it on the slowly moving electrons in the conduction band and on the slowly moving holes in the valence band giving rise to their Landau quantizations. This representation does not lead to the doubling of the magnetic field strength B in a given point. Everywhere it is the same giving rise to the Zeeman effects inside the atomic electron shells and influencing the slowly moving electrons and holes.

In this representation we can write

$$\left(\vec{p}_{\vec{r}} - \frac{q}{c}\vec{A}(\vec{r})\right)^2 \psi(\vec{\rho},\vec{R}) = \left[\vec{p}_\rho - \frac{q}{c}\vec{A}(\vec{\rho}) + \vec{p}_{\vec{R}} - \frac{q}{c}\vec{A}(\vec{R})\right]^2 \psi(\vec{\rho},\vec{R}) =$$
$$= \left\{\left(\vec{p}_\rho - \frac{q}{c}\vec{A}(\vec{\rho})\right)^2 + \left(\vec{p}_{\vec{R}} - \frac{q}{c}\vec{A}(\vec{R})\right)^2 + 2\left(\vec{p}_\rho - \frac{q}{c}\vec{A}(\vec{\rho})\right)\left(\vec{p}_{\vec{R}} - \frac{q}{c}\vec{A}(\vec{R})\right)\right\}\psi(\vec{\rho},\vec{R}) \tag{8}$$

Including the periodic potential $V_0(\rho)$ and taking into account (8), the Pauli equation (4) becomes

$$H\left(\vec{\rho},\vec{R}\right)=H_0\left(\vec{\rho}\right)+\frac{\left(\vec{p}_R-\frac{q}{c}\vec{A}(\vec{R})\right)^2}{2m_0}\hat{I}+\frac{\left(\vec{p}_\rho-\frac{q}{c}\vec{A}(\vec{\rho})\right)\left(\vec{p}_R-\frac{q}{c}\vec{A}(\vec{R})\right)}{m_0}\hat{I};$$

$$H_0\left(\vec{\rho}\right)=\frac{\left(\vec{p}_\rho-\frac{q}{c}\vec{A}(\vec{\rho})\right)^2}{2m_0}\hat{I}+V_0\left(\vec{\rho}\right)\hat{I}-\frac{q\hbar}{2m_0c}\left(\vec{\sigma}\cdot\vec{B}\right) \qquad (9)$$

If the cyclotron frequencies of the conduction electrons and holes are much smaller then the semiconductor energy band gap $E_g$ determined by the periodic potential $V_0(\rho)$, the external magnetic field $\vec{B}=\vec{a}_3 B$ does not change the band structure of the semiconductor. Under this condition the Hamiltonian $H_0(\rho)$ can be transcribed as

$$H_0\left(\vec{\rho}\right)=\left[\frac{\vec{p}_\rho^{\,2}}{2m_0}+V_0(\rho)+\frac{q^2 B^2 \rho^2}{8m_0 c^2}\right]\hat{I}-\frac{qB}{2m_0 c}\left(\hat{I}L_z+2\hat{S}_z\right);\vec{S}=\frac{\hbar}{2}\vec{\sigma};\hat{S}_z=\frac{\hbar}{2}\hat{\sigma}_z. \qquad (10)$$

Here the relations were supposed

$$\vec{A}(\vec{\rho})=\frac{1}{2}\left[\vec{H}\times\vec{\rho}\right],\ \vec{L}(\vec{\rho})=\left[\vec{\rho}\times\vec{p}_\rho\right],$$
$$\text{div}_\rho \vec{A}(\vec{\rho})=0,\ 2\vec{A}(\vec{\rho})\cdot\vec{p}_\rho=\left(\vec{B}\cdot\vec{L}(\vec{\rho})\right). \qquad (11)$$

Operator $\hat{I}\hat{L}_z+2\hat{S}_z$ in (10) can be expressed in the form $\hat{J}_z+\hat{S}_z$, where $\hat{J}_z$ is the projection of the full angular momentum operator $\hat{J}=\hat{L}+\hat{S}$. Its components $J_x, J_y, J_z$ do commute with the Dirac Hamiltonian, whereas the operators $\hat{L}_z$ and $\hat{S}_z$, do not commute separately. It means than the eigenvalues $\hbar j_z$ of the operator $\hat{J}_z$ are the good quantum numbers at least in the absence of the periodic potential $V_0(\rho)$. There are $2j+1$ discrete values of $j_z$, where $j$ determines the eigenvalues of the operator $\hat{\vec{J}}^2$ equal to $\hbar^2 j(j+1)$.

In the presence of the periodic lattice potential $V_0(\vec{\rho})$ the main classification of the energy levels is determined by the irreducible representations of the double crystallographic groups [12]. Nevertheless the quantum numbers $j$ and $j_z$ are needed to describe the Zeeman effects. In atomic physics the operator $\hat{J}_z+\hat{S}_z$ was transformed as follows [11]

$$\left(\hat{J}_z+\hat{S}_z\right)=\hat{G}\hat{J}_z,\ \left(\hat{\vec{J}}+\hat{\vec{S}}\right)=\hat{G}\hat{\vec{J}};$$
$$\hat{G}=1+\frac{\hat{\vec{J}}^2-\hat{\vec{L}}^2+\hat{\vec{S}}^2}{2\cdot\hat{\vec{J}}^2} \qquad (12)$$

The diagonal matrix elements of the operator $\hat{G}$ are obtained using the eigenfunctions of the operators $\hat{J}^2, \hat{L}^2, \hat{S}^2$ and $\hat{J}_z$. They are defined as the atomic g-factors Lande and equal to

$$g = 1 + \frac{j(j+i) - l(l+1) + S(S+1)}{2j(j+1)} \tag{13}$$

Here $l$ and $S$ determine the eigenvalues of the operators $\hat{L}^2$ and $\hat{S}^2$. In the same representation they equal to $\hbar^2 l(l+1)$ and $\hbar^2 S(S+1)$, correspondingly [11].

Now, following the Winkler [12], we will introduce the eigenfunction $U_{n,j,j_z,\vec{k}}(\vec{\rho})$ and eigenvalues $E_{n,j,j_z}(\vec{k})$ of the Hamiltonian $H_0(\vec{\rho})$

$$\hat{H}_0(\rho) U_{n,j,j_z,\vec{k}}(\vec{\rho}) = E_{n,j,j_z}(\vec{k}) U_{n,j,j_z,\vec{k}}(\vec{\rho})$$
$$E_{n,j,j_z}(\vec{k}) = E_{n,j}(\vec{k}) - \frac{q\hbar B}{2m_0 c} g_{n,j} \cdot J_z^{n,j} \tag{14}$$

The g-factors $g_{n,j}$ are introduced for each quantum state of the electron in the periodic lattice potential $V_0(\vec{\rho})$ similarly to the procedure used in the atomic physics [11].

The potential $V_0(\vec{\rho})$ determines the semiconductor energy band structure, which is established much faster than the slowly varying envelope functions. The slowly developing phenomena such as the Landau quantization, the RSOC and the Zeeman effects take place in the frame of the above picture and will be described below in the envelope function approximation proposed in [12]. For this purpose, the solution of (9) containing the variables $\vec{\rho}$ and $\vec{R}$ and describing two types of processes will be taken in adiabatic approximation making its averaging on the quickly establishing crystal lattice states (14).

Below this program is realized for the simplest case of two s-type conduction band electron wave functions with $S_z^c = \pm 1/2$ and of two p-type valence band electron wave functions with $j = +3/2$ and $j_z^V = \pm 3/2$, which are taken in the form

$$\left|\psi_{C,S_z^c}(\vec{P},\vec{R})\right\rangle = U_C(\vec{\rho})\left|S_z^c\right\rangle \varphi_{C,S_z^c}(\vec{R}); S_z^c = \pm\frac{1}{2}$$
$$\left|\psi_{V,M_V,j_z^V}(\vec{\rho},\vec{R})\right\rangle = U_{V,M_V}(\vec{\rho})\left|S_z^V\right\rangle \varphi_{V,j_z^V}(\vec{R}); j_z^V = \pm\frac{3}{2}$$
$$U_{V,M_V}(\vec{\rho}) = \frac{1}{\sqrt{2}}\left(U_{V,p,x}(\vec{\rho}) \pm i U_{V,p,y}(\vec{\rho})\right); M_V = \pm 1 \tag{15}$$
$$j_z^V = M_V + S_z^V$$

To obtain the values $j_z^V = +3/2$, one has to combine the orbital wave function $U_{V,M_V}(\vec{\rho})$ taken with $M_V = 1$ and the spinor-type wave function $\left|S_z^V\right\rangle = \left|\uparrow\right\rangle$ with $S_z^V = 1/2$. In the same way the wave function with $j_z^V = -3/2$ is constructed using the orbital component with $M_V = -1$ and the spinor component with $S_z^V = -1/2$. The spinor components for the electron in the conduction and valence bands $|S_z^c\rangle$ and $|S_z^V\rangle$ coincide ($S_z^c = S_z^V$) for the optical quantum transitions.

In more general form the eigenfunction of the Hamiltonian (9) can be written as

$$\left|\psi(\vec{\rho},\vec{R})\right\rangle = \sum_{S_z^c = \pm 1/2} \left|\psi_{C,S_z^c}(\vec{\rho},\vec{R})\right\rangle + \sum_{j_z^V = \pm 3/2} \left|\psi_{V,M_V,j_z^V}(\vec{\rho},\vec{R})\right\rangle \qquad (16)$$

To obtain the adiabatic approximation, the stationary Schroedinger equation in the form $\hat{H}|\psi(\vec{\rho},\vec{R})\rangle = E|\psi(\vec{\rho},\vec{R})\rangle$ is multiplied successively from the left side by the functions $U_c^*(\vec{\rho}) \times \langle S_z^c|$ and $U_{V,M}^*(\vec{\rho})\langle S_z^V|$ and integrated over the variable $\vec{\rho}$ inside the volume $V_0$ of the elementary lattice cell, taking into account the normalization and orthogonality condition for the periodic functions $U_c(\vec{\rho})$ and $U_{V,M_V}(\vec{\rho})$.

In the adiabatic approximation the equations which determine the slowly varying envelope functions are

$$\begin{aligned}
&\left[E_c + \frac{\left(\hat{\vec{P}}_{\vec{R}} - \frac{q}{c}\vec{A}(\vec{R})\right)^2}{2m_0} - \frac{q\hbar B}{2m_0 c} \cdot g_c S_z^c\right]\varphi_{c,S_z^c}(\vec{R}) + \\
&+ \frac{\left(\hat{\vec{P}}_{\vec{R}} - \frac{q}{c}\vec{A}(\vec{R})\right)}{m_0} \cdot \vec{\pi}_{c;V,M_V}\varphi_{V,j_z}(\vec{R}) = E\varphi_{c,S_z^c}(\vec{R}),\ S_z^c = \pm \tfrac{1}{2}, \\
&\left[E_V + \frac{\left(\hat{\vec{P}}_{\vec{R}} - \frac{q}{c}\vec{A}(\vec{R})\right)^2}{2m_0} - \frac{q\hbar B}{2m_0 c} g_{V,3/2} j_z^V\right]\varphi_{V,j_z^V}(\vec{R}) + \\
&+ \frac{\left(\hat{\vec{P}}_{\vec{R}} - \frac{q}{c}\vec{A}(\vec{R})\right)^2}{m_0} \cdot \vec{\pi}_{V,M_V;c}\varphi_{c,S_z^c}(\vec{R}) = E\varphi_{V,j_z^V},\ j_z^V = \pm \tfrac{3}{2}.
\end{aligned} \qquad (17)$$

Here, the matrix elements are

$$\vec{\pi}_{c,V} = \vec{\pi}_{c;V,M_V} = \vec{\pi}_{V,M_V;c}^* = \frac{1}{V_0}\int_{V_0} d\vec{\rho}\, U_c^*(\vec{\rho})\left(\hat{\vec{P}}_{\vec{\rho}} - \frac{q}{c}\vec{A}(\vec{\rho})\right)U_{V,M_V}(\vec{\rho}) \qquad (18)$$

The equations (14) will be solved when the eigenvalues E are situated in the vicinity of the conduction band bottom and of the valence band top, when the equalities hold

$$E = E_c + \varepsilon_c;\ E = E_V + \varepsilon_V;\ \varepsilon_c, \varepsilon_V \ll E_g^0 = E_c - E_V \qquad (19)$$

Here $E_g^0$ in the semiconductor energy band gap.

In the first case the envelope function $\varphi_{c,S_z^c}(\vec{R})$ plays the main role and according to [12] the auxilliary functions $\varphi_{V,j_z^V}(\vec{R})$ can be determined approximately from the second equation (17) as

$$\varphi_{V,j_z^V}(\vec{R}) = \frac{\left(\hat{\vec{P}}_{\vec{R}} - \frac{q}{c}\vec{A}(\vec{R})\right)}{m_0} \cdot \vec{\pi}_{V,M_V;c} \cdot \varphi_{c,S_z^c}(\vec{R}) \quad (20)$$

The quantum numbers $j_z^V, M_V$ and $S_z^c$ are combined in the way $j_z^V = M_V + S_z^c$ and are realized in the forms $\pm 3/2 = \pm 1 \pm 1/2$. Substituting the functions (20) in the first equation (17) we obtain the terms containing two scalar products of the type $\vec{P}_{\vec{R}} \cdot \vec{\pi}_{V,c}$, which can be approximated by isotropic expressions

$$\left(\hat{\vec{P}}_{\vec{R}} - \frac{q}{c}\vec{A}(\vec{R})\right) \cdot \vec{\pi}_{V,M_V;c} \left(\hat{\vec{P}}_{\vec{R}} - \frac{q}{c}\vec{A}(\vec{R})\right) \cdot \vec{\pi}_{c;V,M_V} \approx$$
$$\approx \left(\hat{\vec{P}}_{\vec{R}} - \frac{q}{c}\vec{A}(\vec{R})\right)^2 \cdot |\vec{\pi}_{c,V}|^2 \quad (21)$$

This procedure simplifies the determination of the electron effective mass. Using this approximations one can write the Schrodinger equation for the envelope wave function $\varphi_{c,S_z^c}(R)$ in the form

$$\left[\frac{1}{2m_e}\left(\hat{\vec{P}}_{\vec{R}} - \frac{q}{c}\vec{A}(\vec{R})\right)^2 - \frac{q\hbar B}{2m_0 c} g_c S_z^c\right]\varphi_{c,S_z^c}(\vec{R}) = \varepsilon_c \varphi_{c,S_z^c}(\vec{R}); \; S_z^c = \pm 1/2 \quad (22)$$

Taking in Eq.(22) the negative charge $q = -e = -|e| < 0$ and the effective mass $m_e$ for the electron, we obtain the final equation

$$\frac{m_0}{m_e} = \frac{2|\vec{\pi}_{cV}|^2}{m_0 \cdot E_g^0} + 1; \; +\frac{|e|\hbar B}{2m_0 c} g_c S_z^c = \mu_B B g_c S_z^c; \; \mu_B = \frac{|e|\hbar}{2m_0 c} \quad (23)$$

For two projections $S_z^c = \pm 1/2$ Eq. (22) can be rewritten in the form of the Pauli equation

$$\left[\frac{\left(\hat{\vec{P}}_{\vec{R}} + \frac{|e|}{c}\vec{A}(\vec{R})\right)^2}{2m_e}\hat{I} + \mu_B B \frac{g_e}{2}\hat{\sigma}_z\right]\hat{\varphi}_e(\vec{R}) = \varepsilon_e \hat{\varphi}_e(\vec{R}),$$
$$\hat{\varphi}_e(\vec{R}) = \begin{vmatrix} \varphi_1 \\ \varphi_2 \end{vmatrix}, \; g_e = g_c. \quad (24)$$

This equation describes the Landau quantization of the conduction electrons together with the spin Zeeman effect and can be supplemented by the term introducing the Rashba spin orbit coupling and will be used below.

In the same way the eigenvalue $E = E_V + \varepsilon_V$ can be investigated using the auxiliary envelope function $\varphi_{c,S_z^c(R)}$ in the form [12]

$$\varphi_{c,S_z^c}(\vec{R}) \cong -\frac{\left(\hat{\vec{P}}_{\vec{R}} - \frac{q}{c}\vec{A}(\vec{R})\right)}{m_0} \cdot \vec{\pi}_{c;V,M_V} \varphi_{V,j_z^V}(\vec{R}), \quad (25)$$

and the same relation between the quantum numbers $j_z^V = S_z^c + M_V$. The Schrödinger equation for the main envelope functions $\varphi_{V,j_z^V}(\vec{R})$ reads

$$\left\{-\frac{1}{2m_h}\left(\hat{\vec{P}}_{\vec{R}} - \frac{q}{c}\vec{A}(\vec{R})\right)^2 - \frac{q\hbar B}{2m_0 c} g_{V,3/2} j_z^V\right\} \varphi_{V,j_z^V}(\vec{R}) = \varepsilon_V \varphi_{V,j_z^V}(\vec{R}), \quad (26)$$

$$j_z^V = \pm 3/2.$$

This equation will be applied to the holes arising in the valence band with the electric charge $q = e = |e| > 0$, the mass $m_h$, the hole energy $\varepsilon_h$ and the hole quantum number $j_z^h$ of the full angular momentum projection as follows

$$\frac{m_0}{m_h} = \frac{2|\vec{\pi}_{c,V}|^2}{m_0 E_g^0} - 1; \ \varepsilon_h = -\varepsilon_V; \ j_z^h = -j_z^V \quad (27)$$

Then Eq. (23) looks as

$$\left\{\frac{1}{2m_h}\left(\hat{\vec{P}}_{\vec{R}} - \frac{|e|}{c}\vec{A}(\vec{R})\right)^2 - \frac{|e|\hbar B}{2m_0 c} g_{V,3/2} j_z^h\right\} \varphi_{V,j_z^V}(\vec{R}) = \varepsilon_h \varphi_{V,j_z^V}(\vec{R}), \quad (28)$$

$$j_z^h, j_z^V = \pm 3/2.$$

Eq. (26) based on the valence band wave functions (15) can be generalized considering also the electron-hole and exciton states with $M_h = j_z^h + S_z^e$ equal to $\pm 2$, as well as the light holes with $j_z^e = \pm 1/2$. In the bulk crystals of the type GaAs the spinor-type wave functions of the heavy and light holes form the four fold degenerate irreducible representation $\Gamma_8$ of the double point symmetry group $T_d$ at the point $\vec{k} = 0$ of the Brillouin zone. In the 2D semiconductor structures such as [001]-grown GaAs/AlAs QWs the symmetry group $T_d$ is reduced to $D_{2d}$ [8] having only two fold degenerate irreducible representations formed by the spinor-type wave functions. They give rise to the second order Pauli matrices instead of the four order matrices in the case of the representations $\Gamma_8$. By this reason the heavy-hole states with $j_z^h = \pm 3/2$ are separated from the light-hole states with $j_z^e = \pm 1/2$ in spite of the fact that they have the same quantum number $j = 3/2$. The separated heavy-hole states with $j_z^h = \pm 3/2$ are denoted by the effective spin $\tilde{S}^h$ with two eigenvalues $\tilde{S}_z^h = \pm 1/2$ correspondingly [8, 13]. Such description is true if the separation between the energy subbands of the heavy and light holes in 2D structures is much greater than any Zeeman splittings or other separate effects [8, 13]. In this case Eq.(28) can be written in the form of the Pauli equation

$$\left\{ \frac{1}{2m_h}\left( \hat{\vec{P}}_{\vec{R}} - \frac{|e|}{c}\vec{A}(\vec{R}) \right)^2 \hat{I} - \mu_B B \frac{g_h}{2}\hat{\sigma}_z \right\} \varphi_h(\vec{R}) = \varepsilon_h \varphi_h(\vec{R}), \qquad (29)$$

where

$$\hat{\varphi}_h(\vec{R}) = \begin{vmatrix} \varphi_{h1} \\ \varphi_{h2} \end{vmatrix}; \; g_h = 3g_{V,3/2} \qquad (30)$$

The heavy hole Zeeman splitting terms in Eqs. (28, 29) can be represented in two alternative forms

$$\begin{aligned} &-\mu_B B g_{V,3/2} j_z^h \text{ with } j_z^h = \pm 3/2 \\ &-\mu_B B g_h \tilde{S}_z^h \text{ with } g_h = g_{V,3/2} \cdot 3, \text{ and } \tilde{S}_z^h = \pm 1/2 \end{aligned} \qquad (31)$$

The first is in the full angular momentum representation whereas the second form corresponds to the effective-spin representation [8, 13].

Eqs.(24) and (29) show that the Zeeman splitting Hamiltonian for 2D electrons and heavy holes can be written in the form

$$H_{e-h} = \mu_B B \left[ g_e S_z^e - g_h \tilde{S}_z^h \right] \qquad (32)$$

These equations explain also why the Landau quantization of the electrons, holes and their cyclotron frequencies depend on their effective masses $m_e$ and $m_h$, while their Zeeman splittings and the Bohr magneton $\mu_B$ are determined by the bare electron mass $m_0$. In the case of light holes the g-factor will be 3 times smaller $g_l = g_{V,3/2} = g_h/3$.

Below the Zeeman splittings of the excitons and trions will be considered following Ref.[8, 13]. We consider the magnetic field being directed along the growth axis. The neutral exciton in its ground 1S state is denoted as $X_S$ and the charged excitons or trions are denoted as $X_S^\pm$. There are four spin configurations for the exciton $X_S$ composed from the electron with $S_z^e = \pm 1/2$ and from the heavy hole with $j_z^h = \pm 3/2$. They are denoted as $|\pm 1/2, \pm 3/2\rangle$. Two of them have the antiparallel spin projections $|1/2, -3/2\rangle$ and $|-1/2, 3/2\rangle$. Their total spin projections $M_h = j_z^h + S_z^e$ and their Zeeman energies $E_z\left(S_z^e, \tilde{S}_z^h\right) = \mu_B B\left[g_e S_z^e - g_h \tilde{S}_z^h\right]$ equal to $M_h = -1$ and $E_z = \mu_B B[g_e + g_h]/2$ in the first case and to $M_h = 1$ and $E_z = -\mu_B B[g_e + g_h]/2$ in the second case. The ZS between these two states equals to $\mu_B B[g_e + g_h]$. There are another two exciton states $|1/2, 3/2\rangle$ and $|-1/2, -3/2\rangle$. They are characterized by the values $M_h = 2$ and $E_z = \mu_B B[g_e - g_h]/2$, and by the values $M_h = -2$ and $E_z = -\mu_B B [g_e - g_h]/2$, respectively. The ZS between the last two states equals to $\mu_B B [g_e - g_h]$. The first two states $|1/2, -3/2\rangle$ and $|-1/2, 3/2\rangle$ are characterized by the circular polarization vectors $\vec{\sigma}_{M_h = \pm 1} = \frac{1}{\sqrt{2}} (\vec{a}_1 \pm i \vec{a}_2)$, where $\vec{a}_1$ and $\vec{a}_2$ are the unit vectors oriented in-plane of the 2D layers [14]. As was pointed in [13], these two exciton states participate in the electric-dipole-allowed optical quantum transitions to the ground state of the crystal due to the radiative recombination of the e-h pair accompanied by the emission of the photon. The exciton state $|1/2, -3/2\rangle$ with $M_h = -1$ and energy $E_{ex}(M_h = -1)$ emits along the z axis (Faraday geometry) the photon with circular polarization $\vec{\sigma}^-$ and energy $E_{ph}(\vec{\sigma}^-)$ whereas the exciton state $|-1/2, 3/2\rangle$ with $M_h = 1$ and energy $E_{ex}(M_h = 1)$ emits the photon with the circular polarization $\vec{\sigma}^+$ and energy $E_{ph}(\vec{\sigma}^+)$. These results described in [13] coincide with the selection rules obtained in the 14] using the Hamiltonian description of the electron-radiation interaction in the 2D e-h system. The selection rules in [14] were expressed in the form of scalar products $(\vec{\sigma}^\pm \bullet \vec{\sigma}^*_{M_h})$. In Faraday geometry the light circular polarization vectors $\vec{\sigma}^\pm$ equal to $(\vec{a}_1 \pm i \vec{a}_2)/\sqrt{2}$. As a result in Faraday geometry the states with $M_h = \pm 1$ give rise to the photon polarizations $\vec{\sigma}^\pm$ respectively. It confirms the result of [13] expressed by the equality

$$\Delta = E_{ex}(M_h = -1) - E_{ex}(M_h = 1) = E_{ph}(\vec{\sigma}^-) - E_{ph}(\vec{\sigma}^+) = g_{ex} \mu_B B,$$
$$g_{ex} = g_e + g_h \tag{33}$$

The trions $X_S^+$ are composed from one electron and two heavy holes (e, h, h) with two spin configurations $|\pm 1/2, 3/2, -3/2\rangle$. Then the e-h pairs $(\pm 1/2, \mp 3/2)$ with $M_h = \mp 1$ recombine by radiating in Faraday geometry the photons with the circular polarizations $\vec{\sigma}^\mp$ and leaving the holes in the spin states $\pm 3/2$ respectively. The exciton, heavy-hole and electron g-factors in type I GaAs/Al$_x$Ga$_{1-x}$As QWs were determined experimentally [9]. All three g-factors vary rapidly in dependence on the QW widths $L_z$ and vanishing for values of $L_z$ between 5 and 12 nm.

The Hamiltonians describing the electron and heavy-hole Landau quantization, Rashba spin-orbit coupling and Zeeman splitting have the forms

$$H_e = \hbar \omega_{ce} \left\{ \left( a^\dagger a + \frac{1}{2} \right) \hat{I} + i\alpha \begin{vmatrix} 0 & a \\ -a^\dagger & 0 \end{vmatrix} + \frac{g_e \mu_B B}{2 \hbar \omega_{ce}} \hat{\sigma}_z \right\},$$

$$H_h = \hbar \omega_{ch} \left\{ \left[ \left( a^\dagger a + \frac{1}{2} \right) + \delta \left( a^\dagger a + \frac{1}{2} \right)^2 \right] \hat{I} + i\beta 2\sqrt{2} \begin{vmatrix} 0 & (a^\dagger)^3 \\ -a^3 & 0 \end{vmatrix} - \frac{g_h \mu_B B}{2 \hbar \omega_{ch}} \hat{\sigma}_z \right\} \tag{34}$$

Here we introduced the Bose-type operators $a^\dagger, a$, which in coordinate representation and acting on the Landau quantization functions are

$$a^\dagger = \frac{1}{\sqrt{2}}\left(\eta - \frac{\partial}{\partial \eta}\right), \quad a = \frac{1}{\sqrt{2}}\left(\eta + \frac{\partial}{\partial \eta}\right), \quad \eta = \frac{y}{l} - pl, \tag{35}$$

where $l$ is the magnetic length and $p$ is the one-dimensional wave number which determines the center of the oscillation.

## 3. Landau quantization, Rashba spin-orbit coupling and Zeeman splitting

Without the Zeeman terms the solutions were obtained in [15-17] and used in [18, 19] for investigation of the optical properties of 2D magnetoexcitons. Here we generalize the solution [16, 17] taking into account the ZS term. We take the spinor type wave function $\hat{\varphi}$ in the form

$$\hat{\varphi} = \begin{vmatrix} f_1 \\ f_2 \end{vmatrix}, \quad f_1 = \sum_n c_n |n\rangle, \quad f_2 = \sum_n d_n |n\rangle,$$
$$\sum_n |c_n|^2 + \sum_n |d_n|^2 = 1 \tag{36}$$

where by $|n\rangle$ are denoted the Fock states with the properties

$$\left(a^\dagger\right)^3 |n\rangle = \sqrt{(n+1)(n+2)(n+3)}\,|n+3\rangle, \quad n \geq 0,$$
$$a^3 |n\rangle = \sqrt{n(n-1)(n-2)}\,|n-3\rangle, \quad n \geq 3 \tag{37}$$

The relations between the coefficients $c_n$ and $d_n$ at $n \geq 3$ are

$$d_{m-3}\left[\left(m - \frac{5}{2}\right) + \delta\left(m - \frac{5}{2}\right)^2 + Z_h - \varepsilon\right] = i\beta 2\sqrt{2}\sqrt{m(m-1)(m-2)}\,c_m,$$
$$c_m\left[\left(m + \frac{1}{2}\right) + \delta\left(m + \frac{1}{2}\right)^2 - Z_h - \varepsilon\right] = -i\beta 2\sqrt{2}\sqrt{m(m-1)(m-2)}\,d_{m-3}, \tag{38}$$
$$m \geq 3$$

Here $\varepsilon$ is the dimensionless eigenvalue and the parameter $Z_h$ determines the magnitude of the ZS in comparison with the cyclotron energy

$$Z_h = \frac{g_h \mu_B B}{2\hbar \omega_{ch}} = \frac{g_h m_h}{4 m_0} \tag{39}$$

The energy spectrum of Landau quantization of the heavy-hole accompanied by the RSOC and by ZS has the form

$$\varepsilon_m^\pm = \frac{E_m^\pm}{\hbar \omega_{ch}} = (m-1) + \frac{\delta}{8}\left[(2m+1)^2 + (2m-5)^2\right] \pm$$
$$\pm \sqrt{\left[\frac{3}{2} - Z_h + \frac{\delta}{8}\left((2m+1)^2 - (2m-5)^2\right)\right]^2 + 8\beta^2 m(m-1)(m-2)}, \tag{40}$$
$$m \geq 3$$

The energy spectrum with quantum numbers $m = 0, 1, 2$ are situated on the energy scale above the solutions $\varepsilon_m^-$ with $m \geq 3$ and are not added here. The only difference between the solutions with and without ZS is the term $Z_h$ in the right hand side of Eq. (40).

Taking into account the RSOC, characterized by the parameter $\beta$, the nonparabolicity of the heavy-hole dispersion law, determined by the parameter $\delta$ as well as the ZS the energy levels $\varepsilon_3^\pm$ are determined by the expression

$$\varepsilon_3^\pm = \frac{E_3^\pm}{\hbar\omega_{ch}} = 2 + 6.25 \cdot 10^{-4} cxy \pm \\ \pm \sqrt{\left[\frac{3}{2} - Z_h + 6 \cdot 10^{-4} cxy\right]^2 + 54.24 \cdot 10^{-4} x^2 y} \tag{41}$$

Here we used the dimensionless units x and y for the external electric and magnetic fields and additional parameters $c$ and $g_h$, which will be used below. This leads to the following expressions for the parameters $\beta$, $\delta$ and $\hbar\omega_{ch}$

$$E_z = x\frac{\text{kV}}{\text{cm}}, \quad B = y\text{T}, \hbar\omega_{ch} = 0.4\text{mev} \cdot y, \quad \delta = 10^{-4} cxy, \quad \beta = 1.062 \cdot 10^{-2} x\sqrt{y}, \quad m_h = 0.25 m_0 \tag{42}$$

Fig.1 shows dependence on the magnetic field strength B of the Zeeman shifts of the 2D heavy hole energy levels $E_3^\pm$ of the Landau quantization in the absence (a) and in the presence (b) of the Rashba spin-orbit coupling. In the absence of the Zeeman effects the solutions are shown by the solid lines, whereas the full solutions $E_3^+$ and $E_3^-$ are shown by the dashed and dotted-dashed lines, correspondingly. In the absence of the RSOC the Landau quantization and the Zeeman splitting effect give rise to the solutions $\varepsilon_3^\pm = 2 \pm \left|\frac{3}{2} - Z_h\right|$, which correspond to the angular momentum projections $j_z^h = \pm 3/2$, respectively. Their splitting $\Delta = \varepsilon_3^+ - \varepsilon_3^-$ equals to $2\left|\frac{3}{2} - Z_h\right|$. At the values $g_h = -48$ and $Z_h = -3$ the splitting $\Delta = 9$ is triple greater in comparison with the case $Z_h = 0$ ($\Delta = 3$), which is clearly seen in the frame (a1) of Fig. 1. The parameters $g_h = 24$ and $Z_h = 3/2$ lead to the degeneracy of two levels $\varepsilon_3^\pm$ and to $\Delta = 0$, shown in the frame (a4) of Fig. 1. Further increase of the Zeeman parameter $Z_h$ from the value 3/2 up to 3 leads to the repelling of two branches and to their returning to the position similar with a small parameter $Z_h$ as it is demonstrated in the frame (a6) of Fig. 1.

In the presence of the RSOC the obtained solutions are drawn in six sections of the group b. The main behavior of the branches remain the same. The difference related with the nonparabolic dispersion law can be observed.

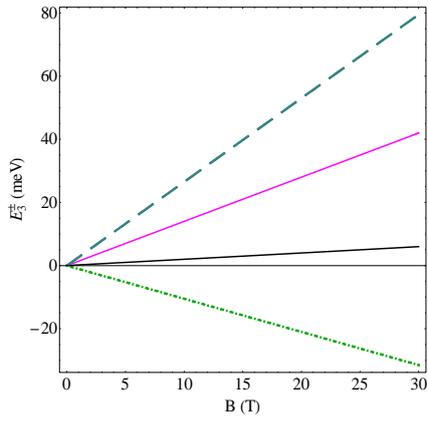
a1) $g_h$=-50, C=0, x=0

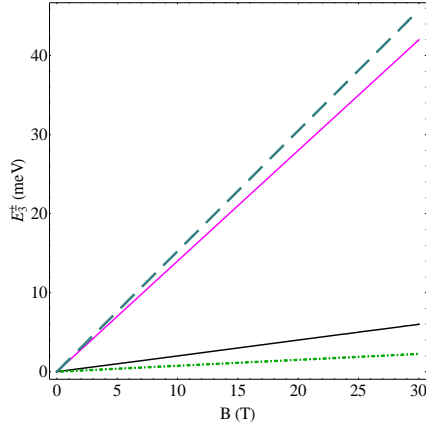
a2) $g_h$=-5, C=0, x=0

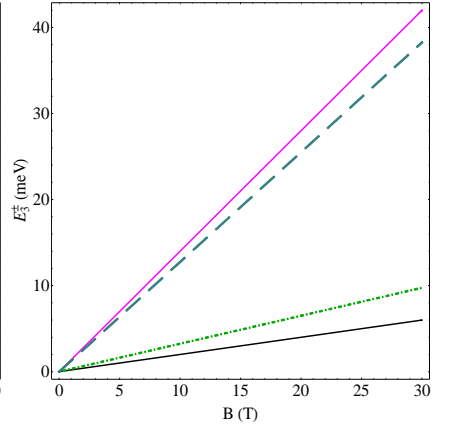
a3) $g_h$=5, C=0, x=0

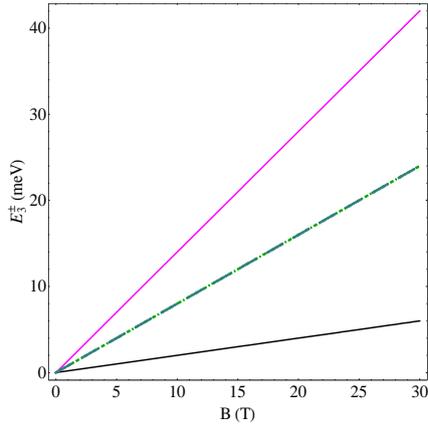
a4) $g_h$=24, C=0, x=0

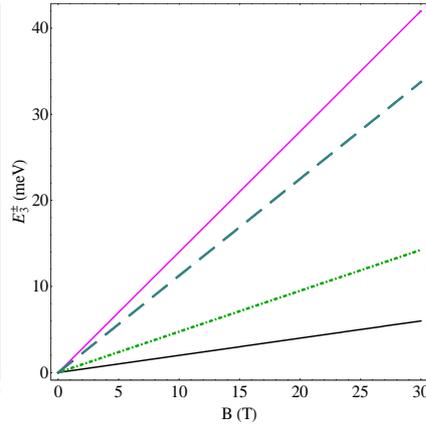
a5) $g_h$=37, C=0, x=0

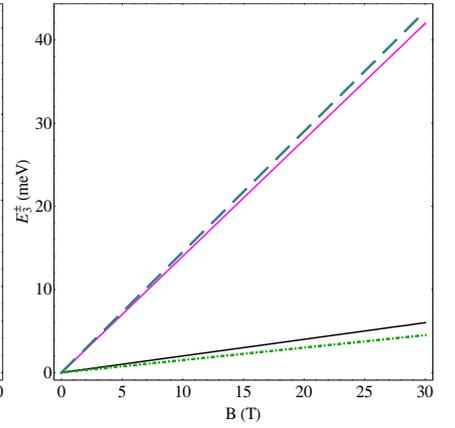
a6) $g_h$=50, C=0, x=0

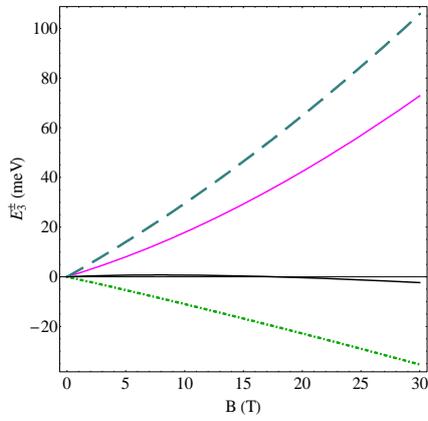
b1) $g_h$=-50, C=5, x=10

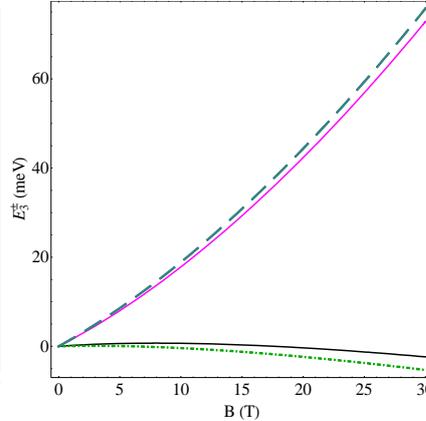
b2) $g_h$=-5, C=5, x=10

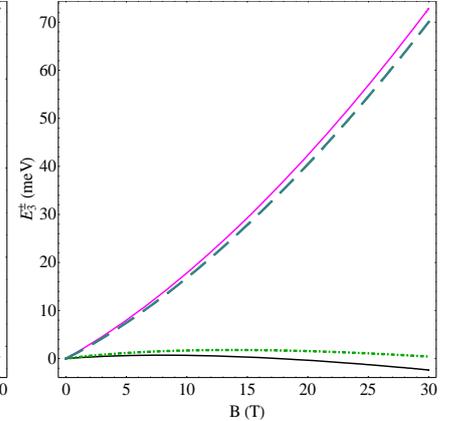
b3) $g_h$=5, C=5, x=10

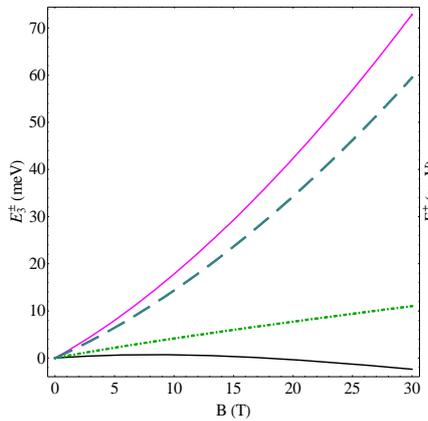
b4) $g_h$=24, C=5, x=10

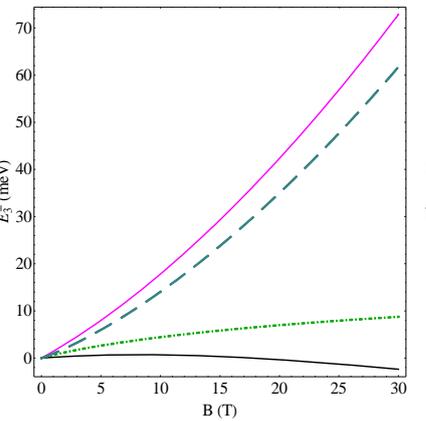
b5) $g_h$=37, C=5, x=10

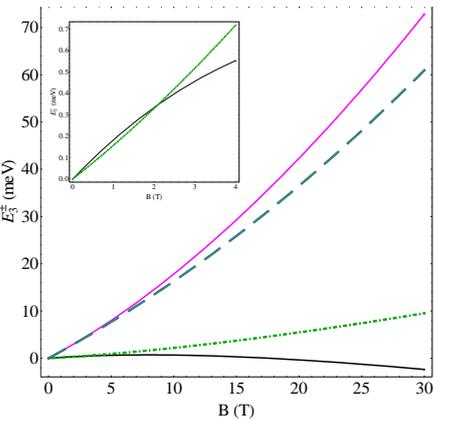
b6) $g_h$=50, C=5, x=10

Fig.1. The Zeeman shifts of the 2D heavy hole energy levels $E_3^\pm$ of the Landau quantization in dependence on the magnetic field strength B at different values of the g-factor. The first group of drawings a1-a6 concerns the absence of the Rashba spin-orbit coupling, whereas the second group b1-b6 takes it into account. The solutions in the absence of the Zeeman effect are represented by the solid lines whereas the full solutions $E_3^+$ and $E_3^-$ by the dashed and dotted-dashed lines. The RSOC and nonparabolic dispersion law are taken into account with the parameters $x = 10$ and $c = 5$.

The main behavior of the pair of Landau quantization energy levels, for example, of the pair $E_3^\pm$, during the increase of the g-factor $g_h$ or equivalently of the Zeeman parameter $Z_h$ proportional to it, when these variables change from negative to positive values consists in the approach of two branches up till a minimal separation on the energy scale followed by their further repel and increasing separation between them. For the parabolic bands the maximal approaches takes place at the value n/2 of the dimensionless parameter $Z_h$, where n coincides with the order of the chirality terms introduced into the Hamiltonian by the RSOC. In our case $n = 3$ and does not depend on the quantum number m in the formula (40). The increasing of the parameter C of the nonparabolicity as well as of the amplitude of the applied perpendicular to the layer electric field increases the separation between the branches.

Fig. 2 depicts the behavior of the energy levels $E_3^\pm$ in dependence on the variable $g_h$, and Fig. 3 shows the combined two previous pictures in the form of two surfaces depending on two variables B and $g_h$. Figs. 4 - 6 show another pairs of the energy levels with the quantum numbers $m = 4,5,6,7$. They confirm the existence of the similar properties as those revealed in the case $m = 3$.

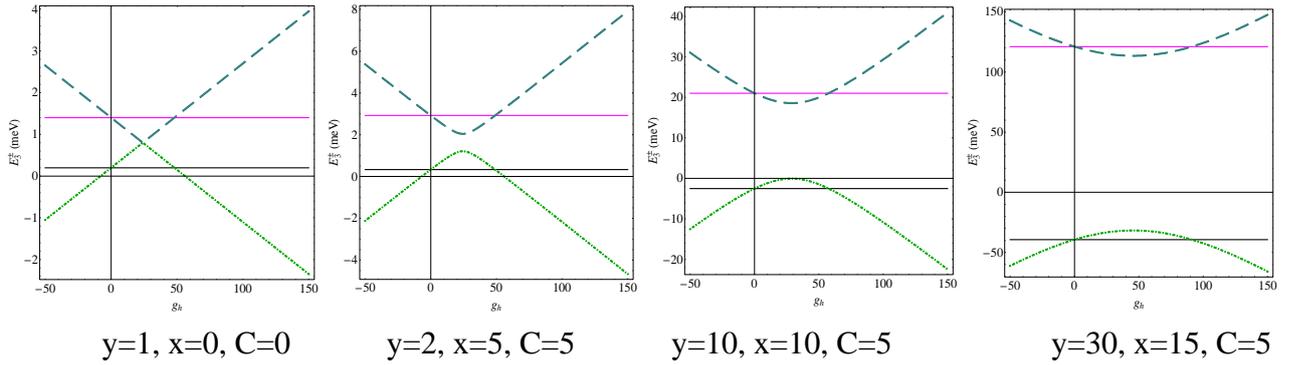

y=1, x=0, C=0     y=2, x=5, C=5     y=10, x=10, C=5     y=30, x=15, C=5

Fig.2. Two branches of the Landau quantization levels $E_3^\pm$ in dependence on the g-factor $g_h$ at different values of the magnetic field strength parameter y, of the RSOC parameter x and of the nonparabolicity parameter C.

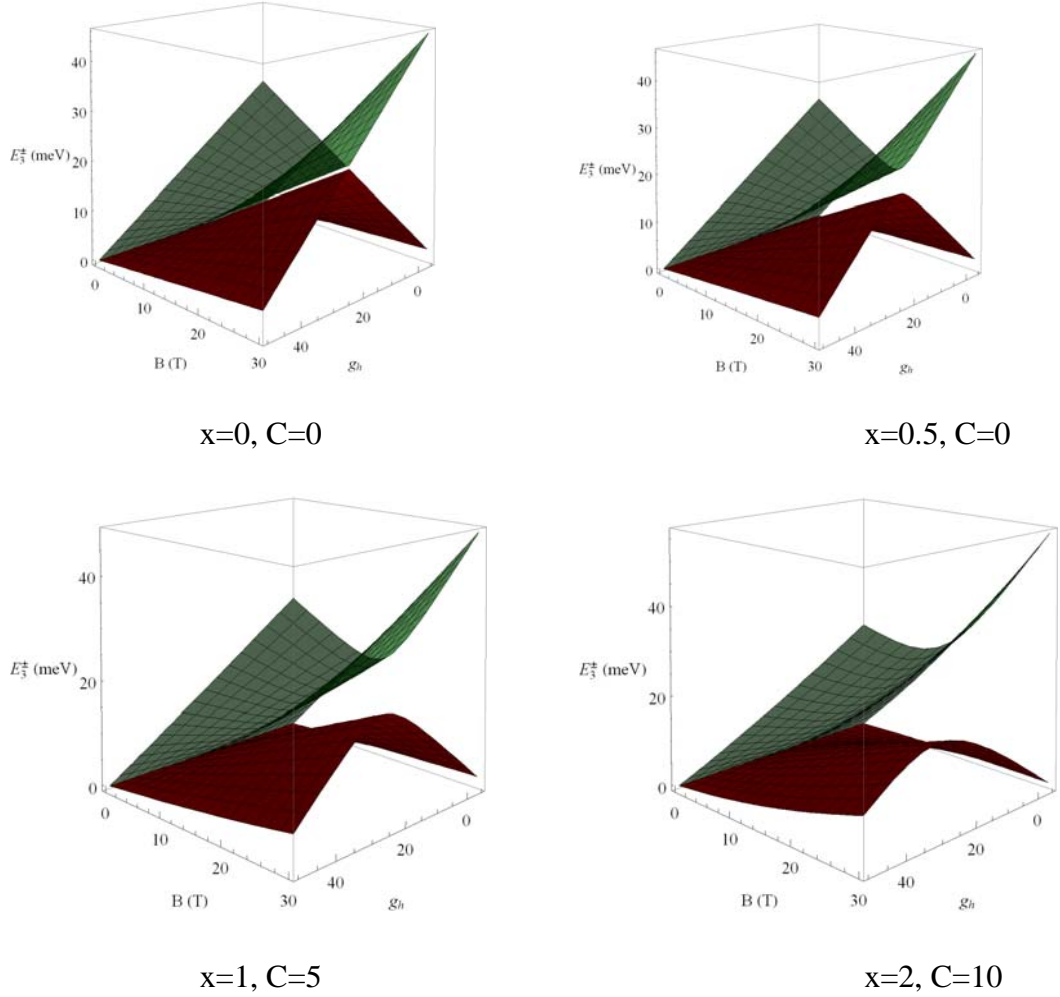

x=0, C=0　　　　　　　　　　　　　　x=0.5, C=0

x=1, C=5　　　　　　　　　　　　　　x=2, C=10

Fig.3. Two surfaces representing the Landau quantization levels $E_3^\pm$ in dependence on the magnetic field strength B, and on the g-factor $g_h$ at different values of the RSOC parameter x and of the nonparabolicity parameter C.

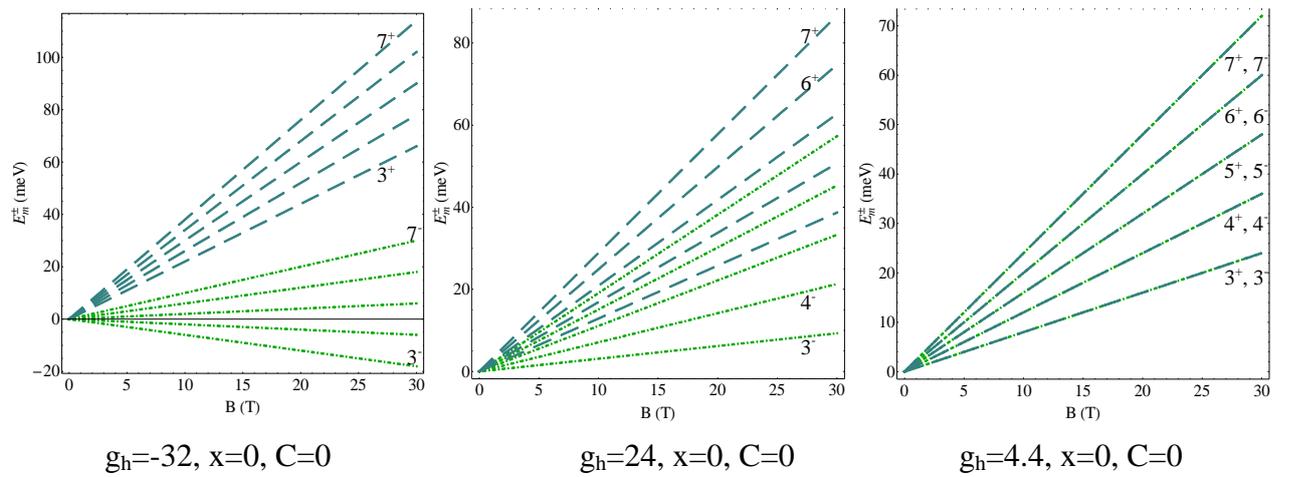

$g_h$=-32, x=0, C=0　　　　$g_h$=24, x=0, C=0　　　　$g_h$=4.4, x=0, C=0

Fig. 4. The dependence of the energy levels $E_m^\pm$ with $m=3,4,5,6,7$ on the magnetic field strength B in the absence of the RSOC, but for different values of the g-factor $g_h$.

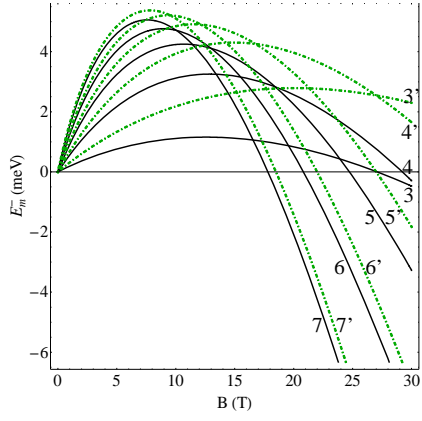 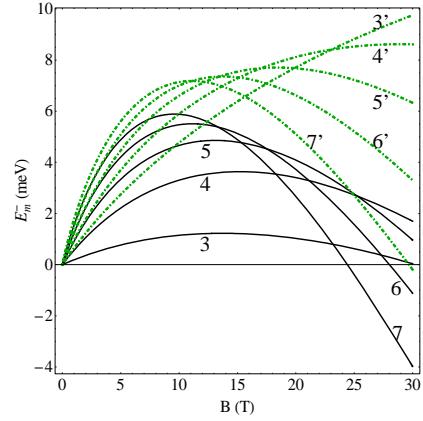

$g_h$=19.8, x=3.8, C=7.38    $g_h$=5, x=3.8, C=5

Fig. 5. The dependence of the energy levels $E_m^-$ with $m = 3,4,5,6,7$ on the magnetic field strength B taking into account the RSOC. The dashed-dotted curves represent the energy levels in the presence of the Zeeman effects, whereas the solid curves without it.

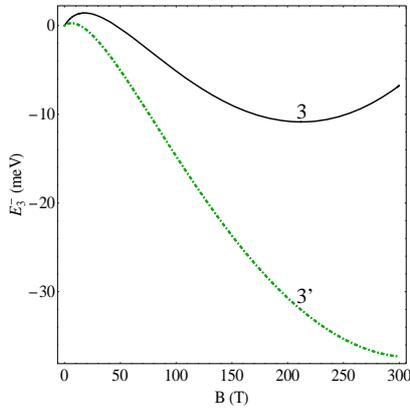 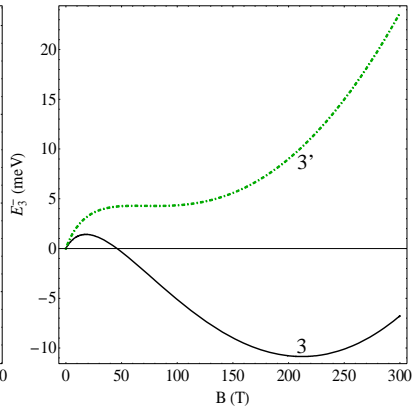

$g_h$=-4.4, x=3.95, C=15    $g_h$=4.4, x=3.95, C=15

Fig. 6. The dependences of the energy levels $E_3^-$ on the magnetic field strength B in the presence of the RSOC and at different values of the g-factor $g_h$. The solid curves 3 correspond to the absence of the Zeeman effect and are drawn for comparison.

**Conclusions**

The origin of the g-factors and effective masses of electrons and holes moving in the crystal lattice potential with the perpendicular magnetic and electric fields was investigated starting with the Dirac equation for the relativistic charged particle with spin 1/2 and finishing with the Pauli equation in the quasi-relativistic approximation. The Pauli equation describing the Landau quantization, the Rashba spin-orbit coupling and the Zeeman splitting for 2D heavy holes with nonparabolic dispersion law induced by the external electric field was solved exactly. The dependences of the energy levels on the magnetic and electric fields strengths and on the g-factor values were discussed.

The main behavior of the pair of Landau quantization energy levels, for example, of the pair $E_3^\pm$, during the increase of the g-factor $g_h$ or equivalently of the Zeeman parameter $Z_h$ proportional to it, when these variables change from negative to positive values consists in the approach of two branches up till a minimal separation on the energy scale followed by their further repel and increasing separation between them. For the parabolic bands the maximal approaches takes place at the value n/2 of the dimensionless parameter $Z_h$, where n coincides with the order of the chirality terms introduced into the Hamiltonian by the RSOC. In the case $n=3$ and does not depend on the quantum number m in the formula (40). The increasing of the parameter C of the nonparabolicity as well as the amplitude of the applied perpendicular to the layer electric field increases the separation between the branches.

**Reference:**
1. E. Brown, Phys. Rev., **133 A**, 1038, (1964)
2. L. Onsager, Phil. Mag., **43**, 1006, (1952)
3. R. Peierls, Z.Physik, **80**, 763, (1933)
4. A. Zee, Quantum Field Theory in a Nutshell, Princeton University Press, 2 edition, (2010) 576 pages
5. L. M.Roth, B. Lax and S. Zwerdling, Phys. Rev., **114**, 90, (1959)
6. J. M. Luttinger and W. Kohn, Phys. Rev. **97**, 869, (1955)
7. F. Nichele et al., Phys. Rev. B, **89**, 081306(R), (2014)
8. H.-W. van Kesteren, E. C. Cosman, W. S. J. A. van der Poel and C. T. Foxon, Phys. Rev. B, **41**, 5283, (1990)
9. M. J. Snelling, E. Blackwood, C. J. McDonagh and R. T. Harley, and C. T. B. Foxon, Phys. Rev. B, **45**, 3922(R), (1992)
10. A. Rahimi - Iman et al., Phys. Rev. B, **84**, 165325, (2011)
11. A. S. Davydov, Quantum mechanics, Pergamon Pr; 2 edition, (1976) 636 pages
12. R. Winkler, Spin-Orbit Coupling Effects in Two-Dimensional Electron and Hole Systems, Berlin, Springer Tracts in Modern Physics, vol 191 (2003)
13. Kyn-Seok Lee, C.D.Lee, S.K.Noh, J. Korean Phys. Soc., **46**, 1410, (2005)
14. S. A. Moskalenko, I. V. Podlesny, M. A. Liberman, B. V. Novikov, J. Nanophotonics, **6**, 061806 (2012)
15. E. I. Rashba, Fiz. Tverd. Tela, **2**, 1224 (1960)
16. S. A. Moskalenko, I. V. Podlesny, P. I. Khadzhi, B. V. Novikov, A. A. Kiselyov, Solid State Commun., **151**, 1690 (2011).

17. T. Hakioglu, M. A. Liberman, S. A. Moskalenko, I. V. Podlesny, J. Phys.: Cond. Matt., **23**, 345405 (2011)
18. I. V. Podlesny, S. A. Moskalenko, T. Hakioglu, A. A. Kiselyov, L. Gherciu, Physica E, **49**, 44 (2013)
19. S. A. Moskalenko, I. V. Podlesny, E. V. Dumanov, A. A. Kiselyov, arXiv:1309.1053 (2013); J. Nanoelectron. and Optoelectron., **9**, 1–20, (2014).